\documentclass[11pt]{article}

\usepackage{amsfonts}
\usepackage{amsmath}
\usepackage{graphicx}
\usepackage{color}
\usepackage{vmargin}
\setpapersize{USletter}
\setmarginsrb{1in}{1.00in}{1in}{0.75in}{0in}{0in}{11pt}{0.4in} % left, top, right, bottom, header, head sep, footer, foot skip
\begin{document}

\title{Computation in Large-Scale Scientific and Internet \\
Data Applications is a Focus of MMDS 2010}

\author{
Michael~W.~Mahoney%
\footnote{Michael Mahoney (\texttt{mmahoney@cs.stanford.edu}) is in the 
Department of Mathematics at Stanford University.  His research interests 
include algorithmic and statistical aspects of large-scale data analysis, 
including randomized algorithms for very large linear algebra problems and 
graph algorithms for analytics on very large informatics graphs.  }
}

\date{}
\maketitle

The 2010 Workshop on Algorithms for Modern Massive Data Sets (MMDS 2010) 
was held at Stanford University, June 15--18.
The goals of MMDS 2010 were 
(1) to explore novel techniques for modeling and analyzing massive, 
high-dimensional, and nonlinearly-structured scientific and Internet data 
sets; 
and (2) to bring together computer scientists, statisticians, applied 
mathematicians, and data analysis practitioners to promote 
cross-fertilization of ideas.
MMDS 2010 followed on the heels of two previous MMDS workshops.
The first, MMDS 2006, addressed the complementary perspectives brought by 
the numerical linear algebra and theoretical computer science communities to 
matrix algorithms in modern informatics applications~\cite{MMDS06summary}; 
and the second, MMDS 2008, explored more generally fundamental algorithmic 
and statistical challenges in modern large-scale data 
analysis~\cite{MMDS08_arxiv}. 

The MMDS 2010 program drew well over $200$ participants, with $40$ talks 
and $13$ poster presentations from a wide spectrum of researchers in modern 
large-scale data analysis.
This included both academic researchers as well as a wide spectrum of 
industrial practitioners.
As with the previous meetings, MMDS 2010 generated intense interdisciplinary 
interest and was extremely successful, clearly indicating the desire among 
many research communities to begin to distill out and establish the 
algorithmic and statistical basis for the analysis of complex large-scale 
data sets, as well as the desire to move increasingly-sophisticated 
theoretical ideas to the solution of practical problems.

\section*{Several Recurring Themes}

Several themes---recurring melodies, as one participant later blogged, that 
played as background music throughout many of the presentations---emerged 
over the course of the four days of the meeting. 
One major theme was that many modern data sets of practical interest are 
better-described by (typically sparse and poorly-structured) graphs or 
matrices than as dense flat tables.
While this may be obvious to some---after all, both graphs and matrices are 
mathematical structures that provide a ``sweep spot'' between more 
descriptive flexibility and better computational tractability---this also 
poses considerable research and implementational challenges, given the way 
that databases have historically been constructed and the way that 
supercomputers have historically been designed. 
A second major theme was that computations involving massive data are 
closely tied to hardware considerations in ways that are very different 
than have been encountered historically in scientific computing and 
computer science---and this is true both for computations involving a 
single machine (recall recent developments in multicore computing) and for 
computations run across many machines (such as in large distributed data 
centers).

Given that these and other themes were touched upon from many complementary 
perspectives and that there was a wide range of backgrounds among the 
participants, 
MMDS 2010 was organized loosely around six hour-long tutorial presentations.

\section*{Large-Scale Informatics: Problems, Methods, and Models}

On the first day of the workshop, participants heard two tutorials that 
addressed computational issues in large-scale data analysis from two very 
different perspectives.
The first was by Peter Norvig of Google, and the second was by John Gilbert 
of the University of California at Santa~Barbara.

Norvig kicked-off the meeting with a tutorial on ``Internet-Scale Data 
Analysis,'' during which he described the practical problems of running, as 
well as the enormous potential of having, a data center so massive that 
``six-sigma'' events, like cosmic rays, drunken hunters, blasphemous 
infidels, and shark attacks, are legitimate concerns.
At this size scale, the data can easily consist of billions to trillions of 
examples, each of which is described by millions to billions of features.
In most data-intensive Internet applications, the peak performance of a 
machine is less important than the price-performance ratio. 
Thus, at this size scale, computations are typically performed on clusters 
of tens or hundreds of thousands of relatively-inexpensive commodity-grade 
CPUs, carefully organized into hierarchies of servers, racks, and 
warehouses, with high-speed connections between different machines at 
different levels of the hierarchy.
Given this cluster design, working within a software framework like 
MapReduce that provides stateless, distributed, and parallel computation 
has benefits; developing methods to maximize energy efficiency is 
increasingly-important; and developing software protocols to handle 
ever-present hardware faults and failures is a must.

Given all of this infrastructure, one can then do impressive things, as 
large Internet companies such as Google have demonstrated.
Norvig surveyed a range of applications such as modelling flu trends with 
search terms, image analysis for scene completion (removing undesirable 
parts of an image and filling in the background with pixels taken from a 
large corpus of other images), and using simple models of text to perform 
spelling correction.
In these and other Web-scale applications, simpler models trained on more 
data can often beat more complex models trained on less data.
This can be surprising for those with experience in small-scale machine 
learning, where the curse of dimensionality and overfitting the data are 
paramount issues.
In Internet-scale data analysis, though, more data mean different data, and 
throwing away even rare events can be a bad idea since much Web data 
consists of individually rare but collectively frequent events.

John Gilbert then provided a complementary perspective in his tutorial 
``Combinatorial Scientific Computing: Experience and Challenges.'' 
Combinatorial Scientific Computing (CSC) is a research area at the interface 
between scientific computing and algorithmic computer science; and an 
important goal of CSC is the development, application, and analysis of 
combinatorial algorithms to enable scientific and engineering computations.
As an example, consider so-called fill-reducing matrix factorizations that 
arise in the solution of sparse linear systems, a workhorse for traditional 
high-performance scientific computation.
``Fill'' refers to the introduction of new non-zero entries into a factor, 
and an important component of sparse matrix solvers is an algorithm that 
attempts to solve the combinatorial problem of choosing an optimal ordering
of the columns and rows of the initial matrix in order to minimize the fill.
Similar combinatorial problems arise in scientific problems as diverse as 
mesh generation, 
iterative methods, climate modeling, 
computational biology, and parallel computing.
Throughout his tutorial, Gilbert focused on two broad challenges---the 
challenge of architecture and algorithms, and the challenge of 
primitives---in applying CSC methods to large-scale data analysis.

The ``challenge of architecture and algorithms'' refers to the nuts and 
bolts of getting high-quality implementations to run rapidly on machines, 
\emph{e.g.}, given architectural constraints imposed by communication and 
memory hierarchy issues or the existence of multiple processing units on a 
single chip.
As an example of the impact of architecture on even simple computations, 
consider the ubiquitous three-loop algorithm for multiplying two 
$n \times n$ matrices, $A$ and $B$: foreach~$i,j,k$,
\[
C(i,j) = A(i,k) * B(k,j)  .
\]
It seems obvious that this algorithm should run in $O(n^3)$ time (and it 
does in the Random Access Model of computation); but empirical results 
demonstrate that the actual scaling on real machines of this na\"{i}ve 
algorithm for matrix multiplication can be closer to $O(n^5)$.
Theoretical results in the Uniform Memory Hierarchy model of computation 
explain this scaling behavior, and it is only more sophisticated BLAS-$3$ 
GEMM and recursive blocked algorithms that take into account memory 
hierarchy issues that run in $O(n^3)$ time.

The ``challenge of primitives'' refers to the need to develop algorithmic 
tools that provide a framework to express concisely a broad scope of 
computations; that allow programming at the appropriate level of 
abstraction; and that are applicable over a wide range of platforms, hiding 
architecture-specific details from the users.
Historically, linear algebra has served as the ``middleware'' of 
scientific computing.
That is, by providing mathematical tools, interactive environments, and 
high-quality software libraries, it has provided an ``impedance match'' 
between the theory of continuous physical modeling and the practice of 
high-performance hardware implementations. 
Although there are deep theoretical connections between linear algebra and 
graph theory, Gilbert noted that it is not clear yet to what extent these 
connections can be exploited practically to create an analogous middleware 
for very large-scale analytics on graphs and other discrete data.
Perhaps some of the functionality that is currently being added onto the 
basic MapReduce framework (and that draws strength from experiences in 
relational database management or high-performance parallel scientific 
computing) will serve this role, but this remains to be seen.

\section*{New Perspectives on Old Approaches to Networked Data}

Although graphs and networks provide a popular way to model large-scale 
data, their use in statistical data analysis has had a long history.
Describing recent developments in a broader historical context was the 
subject of tutorials by 
Peter Bickel of the University of California at Berkeley and 
Sebastiano Vigna of the Universit\`{a} degli Studi di Milano.

In his tutorial on ``Statistical Inference for Networks,'' Bickel described 
a nonparametric statistical framework for the analysis 
of clustering structure in unlabeled networks, as well as for parametric 
network models more generally.
As background, recall the basic Erd\H{o}s-R\'{e}nyi (ER) random graph model:
given~$n$ vertices, connect each pair of vertices with probability~$p$.
If $p \gg \log(n)/n$, such graphs are ``dense'' and fairly regular---due to 
the high-dimensional phenomenon of measure concentration, such graphs are 
fully-connected; they are expanders (\emph{i.e.}, there do not exist any 
good cuts, or partitions, of them into two or more pieces); and the 
empirically-observed degrees are very close to their mean.
On the other hand, for the much less well-studied regime $1/n<p<\log(n)/n$, 
these graphs are very sparse and very irregular---such graphs are not even 
fully-connected; and when considering just the giant component, there are 
many small but deep cuts, and empirically-observed degrees can be much 
larger than their mean.
This lack of large-scale regularity is also seen in ``power law'' 
generalizations of the basic ER model; it's signatures are seen empirically 
in a wide range of very large social and information networks; and it 
renders traditional methods of statistical inference of limited usefulness 
for these very large real-world networks.

Bickel considered a class of models applicable to both the dense/regular 
and sparse/irregular regimes, but for which the assumption of statistical 
exchangeability holds for the nodes.
This exchangeability assumption provides a regularity such that any 
undirected random graph whose vertices are exchangeable can be written as a 
mixture of ``simple'' graphs that can be parametrized by a function 
$h(\cdot,\cdot)$ of two arguments.
Popular stochastic blockmodels are examples of parametric models which 
approximate this class of nonparametric models---the block model with $K$ 
classes is a simple exchangeable graph model, and block models can be used 
to approximate a general function $h$.
In this framework, Bickel considered questions of identifiability and 
consistency; and he showed that, under assumptions such as that the 
expected degree is sufficiently high, it is possible to recover ``ground 
truth'' clusters in this model.

Sebastiano Vigna provided a tutorial on ``Spectral Ranking,'' a general 
umbrella name for techniques that apply the theory of linear functions, 
\emph{e.g.}, eigenvalues and eigenvectors, to matrices that do not 
represent geometric transformations, but instead represent some other kind 
of relationship between entities.
For example, the matrix $M$ may be the adjacency matrix of a graph or 
network, where the entries of $M$ represent some sort of binary relations 
between entities. 
In this case, a common goal is to use this information to obtain a 
meaningful ranking of the entities; and a common difficulty is that the 
matrix $M$ may contain ``contradictory'' information---\emph{e.g.}, 
$i$ likes $j$, and $j$ likes $k$, but $i$ does \emph{not} like $k$; or 
$i$ is better than $j$, $j$ is better than $k$, but $i$ is \emph{not} better 
than~$k$. 

A variant of this was considered by J.R. Seely who, in an effort to rank 
children back in 1949, argued that the rank of a child should be defined 
recursively as the sum of the ranks of the children that like him.
In modern terminology, this led to the computation of a dominant \emph{left} 
eigenvector of $M$ (normalized by row to get a stochastic matrix). 
A dual variant was considered by T.H. Wei who, in 1952, wanted to rank 
sports teams and argued that the score of a team should be related to the 
sum of the scores of the teams it defeated. 
This led to the computation of a dominant \emph{right} eigenvector of $M$ 
(with no normalization).
Since then, numerous domain-specific considerations led researchers to 
propose methods that, in retrospect, are variants of this basic framework.  
For example, in 1953, L. Katz was interested in whether individual $i$ 
endorses or votes for individual $j$, and he argued that the importance of 
$i$ depends on not just the number of voters, but on the number of the 
voters' voters, etc., with a suitable attenuation $\alpha$ at each step.
Since, if $M$ is a zero/one matrix representing a directed graph, the $i,j$ 
entry of $M^k$ contains the number of directed paths from $i$ to $j$, he was 
led to compute $1\sum_{n=0}^{\infty}\alpha^nM^n = 1(I-\alpha M)^{-1}$.
Similarly, in 1965, C.H. Hubbell was interested in a form of clustering 
used by sociologists known as clique identification. 
He argued that on can define a status index $r$ by using the recursive 
equation $r = v + rM$, where $v$ is a ``boundary condition'' or ``initial 
preference,'' and this led him to compute 
$v\sum_{n=0}^{\infty}M^n=v(I-M)^{-1}$.

From this broader perspective, the popular PageRank is the damped spectral 
ranking of the normalized adjacency matrix of the web graph; the boundary 
condition is the so-called preference vector; and this vector can be used 
for various generalizations such as to bias PageRank with respect to a topic
or to generate trust scores.
Remarkably, although PageRank is one of the most talked-about algorithms ever,
there is no reproducible scientific proof that it works for the problem of 
ranking web pages, there is a large body of empirical evidence that it does 
not work, and it is likely to be of miniscule importance in today's ranking 
algorithms.
Nevertheless, partly because the basic ideas underlying spectral ranking are 
so intuitive, there are ``gazillions'' of small variants that could be (and 
are still being) introduced regularly in many areas of machine learning and 
data analysis.
Unfortunately, this is often without reproducible scientific justification 
or careful evaluation of which variants are meaningful or useful.

\section*{Matrix Computations---in Data Applications}

Challenges and tradeoffs in performing matrix computations in MMDS 
applications were the subject of the final pair of tutorials---one by 
Piotr Indyk of the Massachusetts Institute of Technology, and one by
Petros Drineas of Rensselaer Polytechnic Institute.

Indyk discussed recent developments in ``Sparse Recovery Using Sparse 
Matrices.''
This problem arises when the data can be modeled by a vector $x$ that is 
sparse in some (often unknown) basis; and it has received attention 
recently in areas such as compressive sensing, data stream computing, and 
combinatorial group testing.
Traditional approaches first capture the entire signal and then process it 
for compression, transmission, or storage.
Alternatively, one can obtain a succinct approximate representation by 
acquiring a small number of linear measurements of the signal.
That is, if $x$ is an $n$-vector, the representation is $Ax$, for some 
$m \times n$ matrix $A$.
Although typically $m \ll n$, the matrix $A$ can be constructed such that 
one can use a recovery algorithm to obtain a sparse approximation to $x$.
It is often useful (and sometimes crucial) that the measurement matrix $A$ 
be sparse, in that it contains very few non-zero elements per column.
For example, sparsity can be exploited computationally---one can compute the
product $Ax$ very quickly if $A$ is sparse.
Similarly, in data stream processing, the time needed to update the sketch
$Ax$ under an update $\Delta_i$ is proportional to the number of non-zero 
elements in the $i$-th column of $A$.

Indyk described tradeoffs that arise when designing recovery schemes to
satisfy the tricriterion of short sketches, low algorithmic complexity, and 
strong recovery guarantees.
Randomization has proved to be an important computational resource, and thus
a key issue has been to identify properties that hold for very sparse random 
matrices 
and also are sufficient to support 
efficient and accurate recovery algorithms.
A key challenge is that, whereas dense random matrices are fairly 
homogeneous (\emph{e.g.}, 
since measure concentrates
their eigenvalues follow Wigner's semicircle 
law), very sparse random matrices are much less regular.
One can say that a matrix $A$ satisfies the $RIP(p,k,\epsilon)$ property if
\[
||x||_p (1-\epsilon) \le ||Ax||_p \le ||x||_p  
\]
holds for any $k$-sparse vector $x$.
(This generalizes the well-known Restricted Isometry Property from $p=2$ to 
general $p$.) 
Although very sparse matrices cannot satisfy the $RIP(2,k,\epsilon)$ 
property, unless $k$ or $\epsilon$ is rather large, Indyk showed that the 
adjacency matrices of constant-degree expander graphs do satisfy this 
property for $p=1$ and that several previous algorithms generalize to very 
sparse matrices if this condition is satisfied.

In his tutorial on ``Randomized Algorithms in Linear Algebra and Large Data 
Applications,'' Petros Drineas used his work on DNA single-nucleotide 
polymorphisms (SNPs) to illustrate the uses of randomized matrix algorithms 
in 
data analysis.
SNPs are sites in the human genome where a nonnegligible fraction of the 
population has one allele and a nonnegligible fraction has a second allele.
Thus, they are of interest in population genetics and personalized medicine.
In addition, they can be naturally represented
as a $\{-1,0,+1\}$ matrix $A$, where $A_{ij}$ represents whether the $i$-th
individual is homozygous for the major allele, heterozygous, or homozygous 
for the minor allele.

While some SNP data sets are rather small, data consisting of thousands or 
more of individuals typed at hundreds of thousands of SNPs are 
increasingly-common.
Size is an issue since even getting off-the-shelf SVD and QR decomposition 
code to run on dense matrices of size, say, $5000 \times 500,000$ is 
nontrivial on commodity laptops.
The challenge is especially daunting if the computations need to be 
performed thousands of times in the course of a cross-validation experiment.
Perhaps less obvious is the issue of interpretability---even if the data 
clusters well in the span of the top~$k$ ``eigenSNPs,''  these eigenSNPs 
cannot be assayed in the lab and they cannot be easily thought about.
Thus, while eigenvector-based methods for dimensionality reduction are 
popular among data analysts, the geneticists were more interested in the~$k$
actual SNPs that were most important.

Drineas described how to address these two challenges---the ``challenge of 
size'' and the ``challenge of interpretability''---in a unified manner.
He described a randomized approximation algorithm for choosing the best set
of exactly~$k$ columns from an arbitrary matrix.
The key structural insight was to choose columns according to an importance 
sampling distribution proportional to the diagonal elements of the 
projection matrix onto the span of the top~$k$ right singular vectors.
These quantities can be computed exactly by computing a basis for that 
space, or they can be approximated more rapidly with more sophisticated 
methods.
Importantly for interpretability, these quantities are the diagonal elements 
of the so-called ``hat matrix,'' and thus they have a natural interpretation
in terms of statistical leverage and diagnostic regression analysis.
Importantly for size and speed, Hadamard-based random projections 
approximately uniformize these scores, washing out interesting structure and 
providing a basis where simple uniform sampling performs well.
This has led in recent years to fast high-quality numerical implementations 
of these and related randomized algorithms.

\section*{Conclusions and Future Directions}

In addition to these tutorial presentations, MMDS participants heard about 
and discussed a wide range of theoretical and practical issues having to do 
with algorithm development and the challenges of working with modern massive
data sets.
As with previous MMDS meetings, the presentations from all speakers can be 
found at the conference website, \texttt{http://mmds.stanford.edu}; and
as with previous MMDS meetings, participant feedback made it clear that 
there is a lot of interest in MMDS as a developing 
research area at the interface between computer science, statistics, 
applied mathematics, and scientific and Internet data applications.
So keep an eye out for future~MMDSs!

\vspace{0.25in}
\noindent
\textbf{Acknowledgments}

\noindent
I am grateful to the numerous individuals who provided assistance prior to 
and during MMDS 2010;
to my co-organizers Alex Shkolnik, Petros Drineas, Lek-Heng Lim, Gunnar 
Carlsson;
and to each of the speakers, poster presenters, and other participants, 
without whom MMDS 2010 would not have been such a success.

\end{document}